\documentclass[journal]{IEEEtran}

\usepackage{graphicx,amssymb,lineno}
\usepackage{amsmath,amsfonts,amssymb}

\newtheorem{definition}{Definition}
\usepackage{algorithm}
\usepackage{algorithmic}
\usepackage[usenames]{color}
\usepackage{epstopdf}
\usepackage{float}
\usepackage{longtable,booktabs}
\usepackage{cite}

\usepackage{chngpage}
\usepackage{array}

\usepackage{graphicx,graphics,color,epsfig,subfigure,graphpap,rotate}
\usepackage{times, verbatim, subfigure, epsfig, graphicx, latexsym, amsmath}
\usepackage{url}
\usepackage{subfigure}
\begin{document}
%
\title{Coalition Game Based Full-duplex Popular Content Distribution in mmWave Vehicular Networks}

\author{Yibing~Wang,
        Hao~Wu,
        Yong~Niu,~\IEEEmembership{Member,~IEEE,}
        Zhu~Han,~\IEEEmembership{Fellow,~IEEE},
        Bo~Ai,~\IEEEmembership{SeniorMember,~IEEE},
        and Zhangdui~Zhong,~\IEEEmembership{SeniorMember,~IEEE}

\thanks{Copyright (c) 2015 IEEE. Personal use of this material is permitted. However, permission to use this material for any other purposes must be obtained from the IEEE by sending a request to pubs-permissions@ieee.org.}

\thanks{Y. Wang, H. Wu, Y. Niu, B. Ai and Z. Zhong are with the State Key Laboratory of Rail Traffic
 Control and Safety, Beijing Jiaotong University, Beijing 100044, China (E-mails:
 18111034@bjtu.edu.cn, niuy11@163.com).} 
\thanks{Z. Han is with the Department of Electrical and Computer Engineering in the University of Houston, Houston, TX 77004 USA, and also with the Department of Computer Science and Engineering, Kyung Hee University, Seoul, South Korea, 446-701 (E-mail:
zhan2@uh.edu).
} %
\thanks{This study was supported by the National Natural Science Foundation of China Grants 61801016, 61725101, and U1834210; and by the National Key R\&D Program of China under Grants 2016YFE0200900 and 2018YFE0207600; in part by the State Key Lab of Rail Traffic Control and Safety, Beijing Jiaotong University, under Grant RCS2019ZZ005; in part by the funding under Grant 2017RC031; in part by the Fundamental Research Funds for the Central Universities, China, under grant number I20JB0200030; in part by NSF EARS-1839818, CNS1717454, CNS-1731424, and CNS-1702850.
}
}%

\maketitle

\begin{abstract}
The millimeter wave (mmWave) communication has drawn intensive attention with abundant band resources. In this paper, we consider the popular content distribution (PCD) problem in the mmWave vehicular network. In order to offload the communication burden of base stations (BSs), vehicle-to-vehicle (V2V) communication is introduced into the PCD problem to transmit contents between on-board units (OBUs) and improve the transmission efficiency. We propose a full-duplex (FD) cooperative scheme based on coalition formation game, and the utility function is provided based on the maximization of the number of received contents. The contribution of each member in the coalition can be transferable to its individual profit. While maximizing the number of received contents in the fixed time, the cooperative scheme also ensures the individual profit of each OBU in the coalition. We evaluate the proposed scheme by extensive simulations in mmWave vehicular networks. Compared with other existing schemes, the proposed scheme has superior performances on the number of possessed contents and system fairness. Besides, the low complexity of the proposed algorithm is demonstrated by the switch operation number and CPU time.

\end{abstract}


\section{Introduction}\label{S1}
In recent years, millimeter wave (mmWave) has been widely used in many fields, among which the mmWave vehicular network is one of them. In this paper, we consider the popular content distribution (PCD) service in mmWave vehicular networks. Contents distributed by the PCD service may be the high definition map of city, video on demands, real-time traffic information, etc. \cite{ex}. Due to the emergence of a large number of multimedia applications, the massive data growth adds the burden on the network capacity and latency. The conventional cellular model which only relies on stationary base stations (BSs) to send the data cannot satisfy the user demands now. In this case, vehicle-to-vehicle (V2V) communications can help to achieve dependable transmissions and take over the business burden of BSs.

In order to further improve the network capacity and make on-board units (OBUs) of vehicles obtain more contents, full-duplex (FD) communications are utilized in the PCD service. The spectral efficiency of the FD communication can be theoretically doubled due to simultaneous data transmitting and receiving \cite{FD15, FD16, FD17}. In the transmissions with FD communications, the interference of transmission links is caused not only by spectrum reusing but the transmit power of each receiver, that is self interference (SI). For the vehicular network, the high-speed mobility of each vehicle in the network results in the fast-varying channel environment and network topology. Against such a backdrop, it is difficult to determine distribution OBUs and their distribution contents of the PCD based on V2V communications. Besides, the OBU that holds the distribution content may have multiple neighboring OBUs, and which of these neighboring OBUs can receive the distribution content successfully needs to be cautiously determined.

\begin{figure*}[htbp]
  \begin{center}
  \includegraphics[width=16cm]{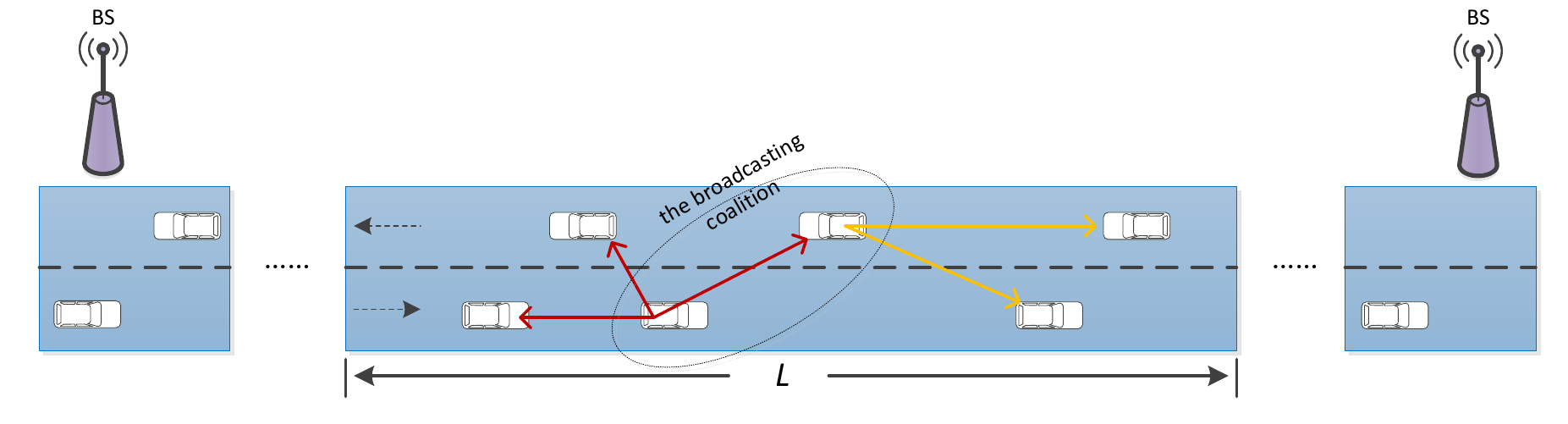}
  \caption{System model of popular content distribution in vehicular ad hoc networks.} \label{fig:highway}
  \end{center}
\end{figure*}

Therefore, we consider a scenario of V2V communications in the mmWave vehicular network. In this paper, all OBUs have a certain amount of initial contents, and they will obtain other required contents by other OBUs broadcast. For the time-varying PCD problem with the high complexity, the coalition formation game can obtain a tractable solution efficiently and directly, and achieve superior performances. Hence, we formulate the problem of how to form broadcasting OBU groups as a coalition formation game. Thus, there are some contributions of this paper, which are mainly summarized as follows
\begin{itemize}
\item We utilize the advantages of mmWave and FD communications for the PCD problem in vehicular networks. And the implementation of V2V communications in the PCD service reduces the pressure of the network. For the proposed cooperation scheme of the mmWave band, OBUs broadcast contents based on the mmWave channel model and directional antenna model. Compared with other schemes, the mmWave and FD communications can obtain higher transmission rates and improve the transmission efficiency significantly.
\item We propose a coalition formation game-based algorithm, which aims at maximizing the number of received contents of all OBUs in the fixed time. The utility function is developed based on the maximization of received content number. Because transmission times of different content transmissions are various, lifetimes of different coalitions are different. The contribution of each coalition member can be transferable to its individual profit, that is the number of neighbors that received contents from the coalition member. The proposed scheme not only maximizes the utility function, but also guarantees the individual profit of each coalition member. Coalitions merge and split according to a series of iterative operations based on the preference relation, and finally self-organize a Nash stable partition. In addition, we analyse the stability and proof the convergence in theory.
\item We propose a broadcasting content selection algorithm to determine contents for OBUs to broadcast, which is based on maximizing the utility function of the coalition formation game-based algorithm. Simulation results show that the proposed scheme makes OBUs receive more contents compared with other schemes. Besides, our scheme has better performance on the fairness of individual profits of broadcasting OBUs. Moreover, we demonstrate that the proposed scheme has the low complexity.

\end{itemize}

The rest of this paper is organized as follows. Section \ref{S2} introduces the related work. In Section \ref{S3}, we provide the system model, including the traffic model, antenna model and channel model. In Section \ref{S4}, we formulate the PCD problem with the objective of maximizing the number of possessed contents. In Section \ref{S5}, a broadcasting content selection algorithm is presented. Section \ref{S6} proposes the coalition formation game-based algorithm, and gives some related definitions and necessary proofs. In Section \ref{S7}, extensive simulations are done to evaluate the proposed scheme. Section \ref{S8} concludes the entire paper.

\section{Related Work}\label{S2}
\begin{table}
  \centering
  \caption{Summary of Notation and Description}\label{Summary}
\begin{adjustwidth}{0.5cm}{0.5cm}
 \begin{tabular}{|p{.2\textwidth}| p{.2\textwidth}| m{.5\textwidth}|}
 \hline
Notation & Description \\
\hline
$M$, $N$, $C$, $L$ & number of TSs, number of OBUs, number of contents, length of the highway \\
\hline
$c$, $D_c$ & content $c$, size of content $c$\\
\hline
$a_i^c$ & binary variable of initial possessed content state of OBU $i$ ($a_i^c=1$ indicates OBU $i$ has received the content $c$ before V2V cooperative transmissions or $a_i^c=0$ indicates OBU $i$ doesn't have content $c$ before V2V cooperative transmissions)\\
\hline
$v_i(0)$ & the initial speed of OBU $i$ in V2V transmission phase \\
\hline
$(i,j)$ & the link from OBU $i$ to OBU $j$  \\
\hline
$P_r(i,j)$ & received power at OBU $j$ from OBU $i$ \\
\hline
$P_t$ & transmit power of links\\
\hline
$G^t_{s_i}$, $G^r_{r_i}$ & directional antenna gain at transmitter $s_i$, directional antenna gain at receiver $r_i$ \\
\hline
$PL(d)$ & path loss for the link of distance $d$ \\
\hline
$D^l$, $\alpha_l$ & path loss of unit length distance, path loss exponent of link $l$\\
\hline
$G_0$, $G_{j,l_{k,r}}$ &  maximum directivity antenna gain, antenna gain at OBU $j$ from link $(k,j)$ \\
\hline
$\beta_j$ &  SI cancelation level at OBU $j$\\
\hline
$I_{i,j}$ &  sum interference of other concurrent links to link $(i,j)$  \\
\hline
$\Lambda_{i,j}$, $c_{i,j}$ &  SINR at the receiver of link $(i,j)$, channel capacity of link $(i,j)$  \\
\hline
$th_{min}$ & SINR threshold of V2V links \\
\hline
$v_{min}$, $v_{max}$, $d_{min}$, $d_{max}$&  lower bound of velocity, upper bound of velocity, lower bound of distance, upper bound of distance \\
\hline
$\theta_{(ij,k)}$, $\theta_{3dB}$, $A_m$ & angle between beam direction of link $(i,j)$ and the direction of OBU $k$, 3dB beamwidth in degrees, maximum attenuation \\
\hline
$\delta_i^c$ & binary variable of the content state of OBU $i$ ($\delta_i^c=1$ indicates OBU $i$ has received the content $c$ or $\delta_i^c=0$ indicates OBU $i$ doesn't have the content $c$ )\\
\hline
$\varphi_{i,j,c}^t$ &  binary variable of the content $c$ ($\varphi_{i,j,c}^t=1$ indicates the content $c$ is transmitted from OBU $i$ to $j$ in slot $t$ or $\varphi_{i,j,c}^t=0$ indicates the content $c$ isn't transmitted from OBU $i$ to $j$ in slot $t$ ) \\
\hline
$B_t$&  number of transmitting antennas of each OBU\\
\hline
$\Gamma_i$, $\mathcal{N}_i$, $\mathcal{N}_i^\ast$ &  possessed content set of OBU $i$, neighbor set of OBU $i$,  set of the neighbors that is sure to receive broadcasting content from OBU $i$\\
\hline
$\gamma_i^p$, $\gamma_{\text{fin}}^i$ &  possessed content of OBU $i$, selected broadcasting content of OBU $i$\\
\hline
$\mathcal{N}_{i^\ast}^{\gamma_i^p}$ &   set of the neighbors of OBU $i$ that need content $\gamma_i^p$ \\
\hline
$\Delta t$, $\alpha$, $\mu$ &   time duration of one slot,  utility calculation factor, pricing factor\\
\hline
 \end{tabular}
\end{adjustwidth}
\end{table}

There are many relevant literatures on V2R or V2V communications. In \cite{H2}, a scheme with low complexity is proposed to schedule packets of downlink and uplink transmissions from one RSU to multiple OBUs. In \cite{H3}, the authors proposed a V2V communication protocol to improve the safety of vehicular networks in the ad hoc network. In \cite{20191}, a V2V-based scheme which jointly optimize the power allocation and modulation/coding is proposed to satisfy requirements of latency and reliability. In \cite{Dilse}, a dependable lattice-based data dissemination scheme for vehicle social internet is presented.

There are also some work on the PCD problem and coalition formation game-based schemes of vehicular networks. In \cite{Deep}, a content centric data dissemination approach based on deep learning is proposed, which considers the vehicle mobility and content type. In \cite{TITS}, a cooperative approach is proposed to achieve dependable content distribution in vehicular networks, which is based on the coalition formation game and big data vehicle trajectory prediction. In \cite{Adcas}, a content popularity-based adaptive caching mechanism is presented to minimize the total operational cost. In \cite{PCD1}, a transmission scheme based on V2V-aided and the coalitional game is proposed to complete the PCD in VANETs. In \cite{zhu}, the authors proposed a cooperation scheme based coalition formation game for RSUs to improve the diversity of the transmission data. In \cite{tian}, the dynamic PCD of vehicular networks is addressed by a proposed cooperation half-duplex (HD) approach based on coalition formation game. The proposed schemes in \cite{zhu} and \cite{tian} aim at minimizing the cost and delay. In our paper, the objective and the utility function of the coalition formation game are both different. We propose a coalition formation game-based algorithm, which aims at maximizing the number of received contents of all OBUs in the fixed time. In addition, \cite{zhu} and \cite{tian} does not consider the performance on fairness of their proposed scheme or evaluate the CPU time of the coalition formation game-based algorithm. In \cite{game}, the authors took advantage of coalition formation game to address bandwidth sharing in V2R communications.

In addition, the research on the problem of the mmWave networks is also a hot spot.
In \cite{add1}, a multihop path selection algorithm and an efficient scheduling scheme for popular content downloading in mmWave small cells are proposed to improve network transmission efficiency.
In \cite{20192}, the authors characterize the interference from the side lanes in mmWave and low terahertz bands. Both multipath interference and direct interference from the transmitting vehicles on the side lanes are considered in \cite{20192}.
In \cite{20193}, a broadcast scheduling scheme with utilizing exclusive region is proposed for improving the concurrency of mmWave links.
In \cite{20194}, the authors proposed a mmWave vehicular framework based on the information-centric network (ICN) to decrease the latency of content dissemination.

However, these existing studies do not combine the mmWave and V2V communications to solve the PCD problem in vehicular networks. And they don't consider how to maximize the number of received contents of all OBUs in the fixed time. In the pursuit of utility function maximization, there is no guarantee for individual profits of OBUs in the existing literature. Hence, we consider the PCD problem in mmWave vehicular networks in this paper. Besides, our current work supports FD communications to improve transmission efficiency. And we propose the new coalition formation game-based scheme which is aimed at maximizing the number of received contents of OBUs in the fixed time.
The proposed scheme considers the channel capacity, content request, peer location, and potential interference to determine the transmission scheduling and obtain superior performances.

\section{System Model}\label{S3}

The system model of the mmWave vehicular network is shown in Fig. \ref{fig:highway}, which contains two lanes with opposite driving directions. There are BSs on both the left and right sides in Fig. \ref{fig:highway}, and different colored arrows represent different transmission contents. OBUs possess some contents that received from BSs at the time of beginning, and then cooperations with other OBUs based on V2V communications provide opportunities for them to obtain other required contents. There is a straight highway of length $L$ in the middle of the Fig. \ref{fig:highway}, and directions of dotted arrows on its left indicate driving directions of vehicles in different lanes. In this paper, we only consider related issues of vehicles sharing content through V2V communications in this highway of length $L$.

On this part of the road, $N$ OBUs are randomly distributed in two lanes, and the set of OBUs is denoted by $\mathcal{N}$. OBUs exchange their possessed contents with other OBUs to obtain as many required files as possible. All OBUs operate in the FD mode, and are equipped with electronically steerable directional antennas to transmit in narrow beams towards other OBUs. Besides, there are $C$ contents in the vehicular network. We assume that OBUs need to acquire these contents, and the set of contents is $\mathcal{C}$. The sizes of different required  contents are different. For the required content $c\in \mathcal{C}$, its size is denoted by $D_c$. At the beginning of the cooperative transmission, each OBU has a random number of contents, which are received from BSs.

For clarity of illustration, the transmission time is divided into a series of non-overlapping time slots (TSs). There are $M$ equal TSs in all for content transmissions. In the following content, the number of TSs is used to measure the length of the transmission time.

Next, the channel model and traffic model of the proposed scheme will be presented in detail.

\subsection{ Channel Model}\label{S3-1}
In the vehicular network, channel model has a significant impact on transmission performances. According to the dynamic channel environments and fast-varying network topologies, we assume the V2V links exist only between neighbor vehicles with line-of-sight (LOS) links, the path loss expressions are given as \cite{3GPP}
\begin{eqnarray}\label{eqPL}
PL(d)=
D^ld^{-\alpha_l},
\end{eqnarray}
where $D^l=(\lambda/{4\pi})^{\alpha_l}$ is constant that represents the path loss of unit length distance of LOS links, and $\lambda $ is the wavelength. $\alpha_l$ is the path loss exponent of LOS which is affected by the environmental scenario. $PL(d)$ represents the path loss for the link of distance $d$.

From a certain recent tractable models \cite{ant2, Na1}, we adopt an independent Nakagami-m distributed channel model for analyzing mmWave channel fading characteristics, where $m$ is the fading depth parameter. The Nakagami-m distributed channel model does not assume the existence of LOS, but uses the gamma function to fit the experimental data. Therefore, the Nakagami-m distributed channel model is more general in mmWave communication. The probability density function (pdf) of the channel envelope $A$ in Nakagami-m distributed is given as \cite{Na2, Na3}
\begin{equation}\label{eqN1}
f_A(a;m,\omega)=\frac{2}{\Gamma(m)}\left(\frac{m}{\omega}\right)^ma^{2m-1}\exp\left(-\frac{m}{\omega}a^2\right),
\end{equation}
where $\omega$ is the average power of the signal, and $\Gamma(m)$ is the gamma function which is defined by $\Gamma(z)=\int_0^\infty x^{z-1}e^{-x}dx$ for any real number $z$$>$0. In the Nakagami-m distributed channel model, the channel power gain is gamma-distributed. The pdf of the channel power gain $H$ in gamma-distributed is given as \cite{Ga1}
\begin{equation}\label{eqH1}
f_H(h;\varrho,\varsigma)=\frac{\varsigma^{-\varrho}h^{\varrho-1}e^{-h/\varsigma}}{\Gamma(\varrho)},
\end{equation}
where $\varrho$ ($\varrho=m$) is the shape parameter, $\varsigma$ ($\varsigma=1/m$) is the scale parameter, and $\Gamma(\varrho)$ is the gamma function with parameter $\varrho$.

For desired link $(i,j)$, the received signal power at OBU $j$ from OBU $i$ can be calculated as
\begin{equation}\label{eqd}
{P_r}(i,j) = P_tG_0h_{i,j}D^ld_{i,j}^{-\alpha_l},
\end{equation}
where $P_t$ is the transmit power of the link, $G_0$ is the maximum directivity antenna gain, and $h_{i,j}$ is the channel power gain of link $(i,j)$ in gamma-distributed (\ref{eqH1}).

For interfering link $(k,j)$, the received signal power at OBU $j$ caused by desired link $(k,r)$ can be calculated as
\begin{equation}\label{eqi}
{P_r}(k,j) = P_tG_{j,l_{k,r}}h_{k,j}D^ld_{k,j}^{-\alpha_l},
\end{equation}
where $G_{j,l_{k,r}}$ is the antenna gain at OBU $j$ from link $(k,j)$ which is calculated as (\ref{eqGi}), and $h_{k,j}$ is the channel power gain of link $(k,j)$ in gamma-distributed channel.

Due to FD communication, self interference (SI) needs to be considered in the V2V phase. For the problem from SI, so many cancellation schemes have been proposed so far, but the existing SI cancellation schemes cannot completely eliminate SI. In fact, SI cancellation is a very complex research focus, so we will not introduce related techniques in the paper. In order to be realistic, we assume that appropriate SI cancellation technology is employed, then the amount of remaining self interference (RSI) can be represented by the transmit power of the link receiver in the calculation. If OBU $j$ are the receiver of link $(i,j)$ and the transmitter of link $(j,f)$ at the same time, the RSI must exist at OBU $j$ and can be denoted by $\beta_jP_{t_{j,f}}$ ($f\in \mathcal{N}$), where $\beta_j$ represents the SI cancelation level at OBU $j$ and is a non-negative parameter. Multiple antennas can send messages simultaneously, so the RSI can come from multiple links at the OBU.

According to equation (\ref{eqd}), (\ref{eqi}) and the expression of RSI, the SINR at the receiver of link $(i,j)$ is given as follows
\begin{equation}\label{eqsi}
\Lambda_{i,j} = \frac{P_tG_0h_{i,j}D^ld_{i,j}^{-\alpha_l}}{N_0W+I_{i,j}+\sum\beta_jP_t},
\end{equation}
where $N_0$ is the onesided power spectral density of white Gaussian noise, $W$ is the channel bandwidth, and $I_{i,j}$ is the sum interference of other concurrent links.
\begin{equation}\label{eqI}
\begin{split}
I_{i,j} = \sum_{k\in\mathcal{N}\setminus\{i\}, r\in\mathcal{N}\setminus\{j\}} P_tG_{j,l_{k,r}}h_{k,j}D^ld_{k,j}^{-\alpha_l}.
\end{split}
\end{equation}

Therefore, the channel capacity of link $(i,j)$ between OBU $i$ and $j$ is given by
\begin{eqnarray}\label{eq5}
c_{i,j}=
Wlog_2\left(1+\Lambda_{i,j}\right).
\end{eqnarray}

In order to guarantee transmission qualities of links, QoS requirements of V2V links are given, that is, the actual SINR of the link must be larger than the given SINR threshold. For all V2V links, $th_{min}$ is defined as the SINR threshold. Therefore, the condition that the V2V link from OBU $i$ to $j$ can be successfully transmitted can be expressed as
\begin{eqnarray}\label{eq6}
\Lambda_{i,j}\geq th_{min}.
\end{eqnarray}



\subsection{ Traffic Model}\label{S3-2}
Considering a mobility model in highway scenarios, which is similar to the Freeway Mobility Model (FMM) in \cite{Model}. At the beginning, all vehicles are randomly distributed in lanes and drive at their initial speeds. To cope with the high-speed movement of the vehicle, we need to obtain state parameters of each vehicle every time slot. Vehicles randomly choose to accelerate or decelerate in every slot.

In order to simplify the network structure, we assume that there is a two-lane freeway without intersections as shown in Fig. \ref{fig:highway}. For two subsequent vehicles in the same lane, the distance between them is limited to $[d_{min}, d_{max}]$. $d_{min}$ is a security distance and $d_{max}$ is an upper bound of distance. In addition to the distance, the velocity of each vehicle is limited to $[v_{min}, v_{max}]$. Thus, the initial speed of OBU $i$ is limited by $v_{min}\leq v_i(0)\leq v_{max}$, and vehicles speed up or slow down by probability $p$. The choice of each vehicle is independent, and the acceleration is $a>0$. We assume that vehicles in each lane will not change the direction of travel on this highway. Limitations of velocity and distance are to make the network model more in line with the actual vehicle network.

According to the above constraints, the speed of OBU $i\in \mathcal{N}$ in slot $t$ satisfies ($0<p<1/2$):
\begin{eqnarray}\label{eq3}
v_i(t+1)=
\begin{cases}
min[v_i(t)+a, v_{max}]\,   &p, \\
max[v_i(t)-a, v_{min}]\,   &p, \\
v_i(t), &1-2p.\\
\end{cases}
\end{eqnarray}

On account of different speeds of different vehicles and limited distances between subsequent vehicles, some vehicles have overtaking behavior in the actual situation. However, there is only single lane of one direction in the setting highway model. In order to ensure the safety of vehicle traveling, we don't consider the situation of the vehicles overtaking in this paper. For any OBU $i\in \mathcal{N}$ with OBU $j$ ahead in the same lane, the constraints of OBU's behavior change are given as follows
\begin{enumerate}
\item If $d_{i,j}\leq d_{min}$, OBU $i$ decelerates to $v_i(t+1)=v_{min}$.
\item If $d_{i,j}\geq d_{max}$, OBU $i$ accelerates to $v_i(t+1)=v_{max}$.
\end{enumerate}

\subsection{ Antenna Pattern}\label{S3-3}
Considering mmWave communications in the vehicular network, directional antenna is necessary to ensure transmission quality \cite{antf}. Thus, what follows is a specific analysis of the antenna model for V2V communications.

We have assumed that all OBUs communicate in the FD mode, which means each vehicle can send and receive messages simultaneously. To simplify the problem, we assume that each vehicle can send messages to multiple vehicles at the same time, but can receive message from only one vehicle. So we adopt multiple antennas to form multiple directional beams at each vehicle for transmitting multiple data links simultaneously \cite{ant0}. The maximum number of beams formed simultaneously at each vehicle is denoted by $B_t$.

\begin{figure}[t!]
  \begin{center}
  \includegraphics[width=8cm]{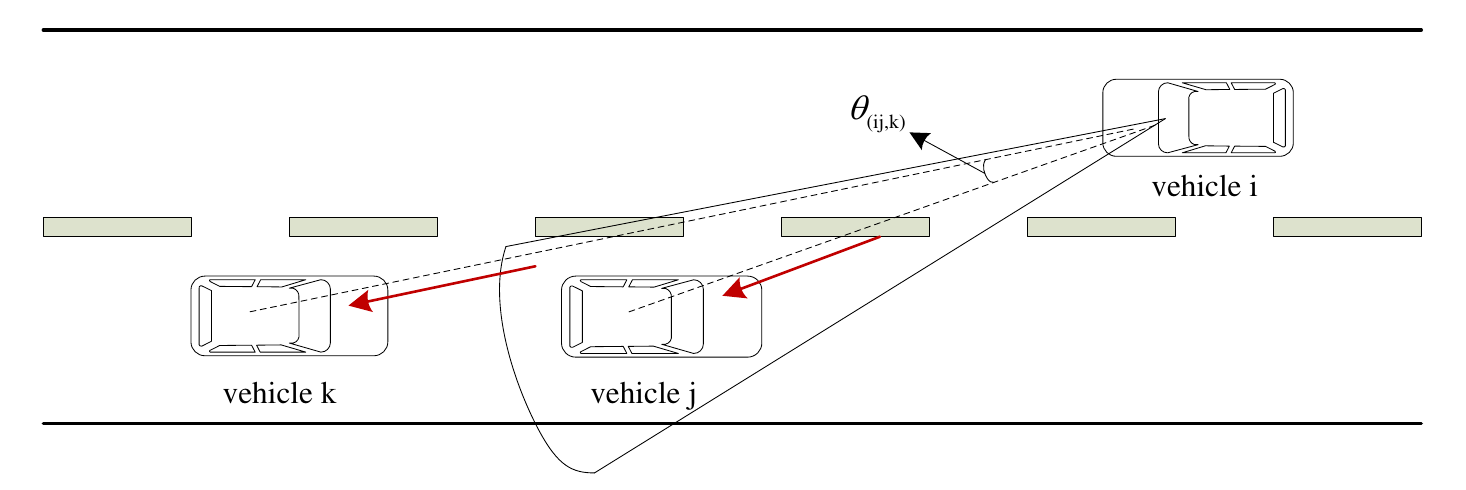}
  \caption{Interference model between V2V communications} \label{fig:antenna}
  \end{center}
\end{figure}

The interference model between V2V communications in the smart antenna is shown as Fig. \ref{fig:antenna}. We assume vehicles have perfect channel state information (CSI) and can steer their antenna orientations for the maximum directivity gain \cite{ant1, ant2}. The perfect beam alignment for the desired link has the maximum directivity antenna gain of $G_0$, i.e., the antenna gain of the link between OBU $i$ and $j$ is $G_0$. We denote the angle between the beam direction and the direction of an OBU by $\theta$, i.e., the angle between beam direction of link $(i,j)$ and the direction of OBU $k$ is $\theta_{(ij,k)}$. Thus, the antenna gain of the interfering link $(i,j)$ at OBU $k$ is specified as \cite{IEEE, V2I}
\begin{equation}\label{eqGi}
G_{k,l_{ij}}=-min\left[12\left(\frac{\theta_{(ij,k)}}{\theta_{3dB}}\right)^2, A_m\right]+G_0, -\pi\leq\theta\leq\pi,
\end{equation}
where $\theta_{3dB}$ is the 3dB beam-width in degrees, and $A_m$ is the maximum attenuation. When $\theta=0$, $G_i=G_0$ is the antenna gain of the desired link.

\section{Problem Formulation}\label{S4}
In this section, we propose the formulation of the cooperative transmission in the vehicular network. This work aims at maximizing the amount of received contents in the fixed time. Moreover, the main challenge is how to jointly determine the broadcasting content of each broadcasting OBU, and how to group broadcasting OBUs from a content number maximization perspective. We define a binary variable $\delta_i^c$ to indicate whether the OBU $i$ has received the content $c$ successfully. If that is the case, we have $\delta_i^c=1$; otherwise, $\delta_i^c=0$. Hence, the objective function of the proposed scheme is given by
\begin{equation}\label{function}
\text{max}\frac{1}{|\mathcal{N}|}\sum_{i\in \mathcal{N}}\sum_{c\in \mathcal{C}}\delta_i^c.
\end{equation}

For the OBU $i$, we define a binary variable $\varphi_{i,j,c}^t$ to indicate whether the content $c$ is transmitted from OBU $i$ to OBU $j$ in slot $t$. If it is the case, we have $\varphi_{i,j,c}^t=1$; otherwise, $\varphi_{i,j,c}^t=0$. Due to the transmission characteristic of the FD mode, each OBU can be the transmitter and receiver at the same time. However, one OBU cannot be the receivers of multiple concurrent links owing to the setting of single receiving antenna of each OBU. Then, we can obtain the related constraint as follows.
\begin{equation}\label{constraint1}
\sum_{i\in\mathcal{N}}\sum_{c\in\mathcal{C}}\varphi_{i,j,c}^t\leq 1, \forall j\in\mathcal{N}.
\end{equation}

In any slot, the broadcast contents of one OBU to other OBUs should be the same. Therefore, we can obtain the following constraint.
\begin{equation}\label{constraint2}
\varphi_{i,j,c}^t+\varphi_{i,j^\prime,c^\prime}^t\leq 1, \forall i\in\mathcal{N}, \forall j,j^\prime\in\mathcal{N}, \forall c,c^\prime\in\mathcal{C}.
\end{equation}

According to the setting of $B_t$ receiving antennas of each OBU, the constraint of OBU $i$ can be given by
\begin{equation}\label{constraint3}
\sum_{j\in\mathcal{N}}\varphi_{i,j,c}^t\leq B_t, \forall i\in\mathcal{N}, \forall c\in\mathcal{C}.
\end{equation}

In summary, the problem of optimal selection and scheduling can be formulated as follows.
\begin{equation}\nonumber
\begin{split}
\text{max}  & \;\; \frac{1}{|\mathcal{N}|}\sum_{i\in \mathcal{N}}\sum_{c\in \mathcal{C}}\delta_i^c.\\
\text{s.t.} & \;\;
(\ref{constraint1})-(\ref{constraint3}).
\end{split}
\end{equation}
This optimal problem is a nonlinear integer programming problem, and is NP-hard \cite{Wang}. We propose the selection approach to maximize the number of received contents in next section. Besides, we will also propose a game theory model and define the utility function of the coalitional game.

\section{Broadcasting Content Selection Algorithm}\label{S5}

Focusing on the V2V cooperative transmission, we first determine the broadcasting content of each OBU. The numbers of slots for different content transmissions are unequal, but the broadcasting content selections of OBUs are all simultaneously executed before the transmissions. In a certain slot $t$, we select the broadcasting contents of OBUs. Due to the short duration of the slot, we assume vehicles are stationary in a time slot. The possessed content set of OBU $i\in \mathcal{N}$ is denoted by $\Gamma_i$. The neighbor set of OBU $i$ is denoted by $\mathcal{N}_i$, and OBUs in $\mathcal{N}_i$ satisfy (\ref{eq6}). The set of the neighbors that is sure to receive broadcasting content from OBU $i$ is denoted by $\mathcal{N}_i^*\subseteq\mathcal{N}_i$. We have assumed the number of transmitting antennas of each OBU is $B_t$, so we have $|\mathcal{N}_i^*|\leq B_t$.
Moreover, the selected broadcasting content of OBU $i$ is denoted by $\gamma_\text{fin}^i$.

For any OBU in the system, it can be the neighbor of several OBUs, but it only can receive the content from one OBU in each slot. If there are LOS links between OBU $i$ and both OBU $m$ and OBU $n$, we can obtain the constraint as follows.
\begin{equation}\label{eq10}
\begin{split}
&i\in\mathcal{N}_m\&i\in\mathcal{N}_n,   \\
&\text{If}\, i\in\mathcal{N}_m^*, \, i\not\in\mathcal{N}_n^*, \\
&\text{If}\, i\in\mathcal{N}_n^*, \, i\not\in\mathcal{N}_m^*.
\end{split}
\end{equation}

OBUs broadcasting different contents has a significant impact on transmission performances. Therefore, we propose a broadcasting content selection algorithm. The goal of the selection algorithm is to successfully transmit broadcasting contents from broadcasting OBUs to as many neighbor OBUs as possible.

First, each broadcasting OBU must select its broadcasting content from its possessed contents. We denote one of the possessed contents of OBU $i$ by $\gamma_i^p\in\Gamma_i$. And then the set of the neighbors that need content $\gamma_i^p$ can be denoted by $\mathcal{N}_{i^*}^{\gamma_i^p}$. When $|{N}_{i^*}^{\gamma_i^p}|$ isn't larger than the number of transmitting antennas of broadcasting OBU, we only compare the value of $|{N}_{i^*}^{\gamma_i^p}|$ to select the broadcasting content. For maximizing the number of OBUs that receive the broadcasting content, $\gamma_i^p$ is selected to be broadcasted by OBU $i$, if and only if
\begin{equation}\label{eq8}
|\mathcal{N}_{i^*}^{\gamma_i^p}|\geq|\mathcal{N}_{i^*}^{\gamma_i^q}|, \,\forall \gamma_i^q\in\Gamma_i, |\mathcal{N}_{i^*}^{\gamma_i^p}|\leq B_t, |\mathcal{N}_{i^*}^{\gamma_i^q}|\leq B_t.
\end{equation}

However, there may be more than $B_t$ neighbors of broadcasting OBUs. According to different content sizes and different transmission rates, we define a parameter to estimate the TS number of content $\gamma_i^p$ transmission from broadcasting OBU $i$ to its neighbor $j$
\begin{equation}\label{TS number}
T_{i,j}^{\gamma_i^p}=\frac{D_{\gamma_i^p}}{c_{i,j}\cdot\Delta t},
\end{equation}
where $c_{i,j}$ is the transmission rate calculated by (\ref{eq5}), and $\Delta t$ is the time duration of one slot. $D_{\gamma_i^p}$ is the size of content $\gamma_i^p$. Since concurrent links are not determined, we do not consider interferences of other links here.

For any possessed content $\gamma_i^p$ of OBU $i$, we select $B_t$ neighbor OBUs $j_1, j_2, ..., j_{|B_t|}\in\mathcal{N}_{i^*}^{\gamma_i^p}$ with relatively small $T_{i,j_1}^{\gamma_i^p}$, $T_{i,j_2}^{\gamma_i^p}$, ..., $T_{i,j_{|B_t|}}^{\gamma_i^p}$ to receive it. In this case, $\gamma_i^p$ is selected to be broadcasted by OBU $i$, if and only if
\begin{equation}\label{eqcom}
\begin{split}
\text{max}\{T_{i,j_1}^{\gamma_i^p}, T_{i,j_2}^{\gamma_i^p}, ..., T_{i,j_{|B_t|}}^{\gamma_i^p} \}\,\leq \text{max}\{T_{i,j^\prime_1}^{\gamma_i^q}, T_{i,j^\prime_2}^{\gamma_i^q}, ..., T_{i,j^\prime_{|B_t|}}^{\gamma_i^q}\}, \\
\forall \gamma_i^q\in\Gamma_i,\; j^\prime_1,j^\prime_2, ...,j^\prime_{|B_t|}\in\mathcal{N}_{i^*}^{\gamma_i^q}.
\end{split}
\end{equation}
Transmitting contents that need less slots preferentially can complete their transmissions quickly and leave slots for other contents. Therefore, OBUs can obtain more contents. If the number of neighbors that need different broadcasting contents are the same, this selection method is equally applicable. The broadcasting content selection algorithm is summarized in Algorithm \ref{alg:The selection algorithm}. In Lines 5-13, selecting the broadcasting content for the broadcasting OBU with more than $B_t$ neighbors according to (\ref{eqcom}). In Lines 17-24, the algorithm does the content selection under the condition of equal number of neighbor OBUs. Finally, we obtain the selected broadcasting content of each OBU and the neighbors that receive this selected content in Line 27.

For the complexity of the selection algorithm, the outer {\tt for} loop has $\mathcal{O}(N)$ iterations. The number
of the iterations of the {\tt for} loop in Line 2 is $|\Gamma_i|$,
where $|\Gamma_i|$ in the worst case is $\mathcal{O}(C)$. The {\tt for} loop in
Line 6 has $|\mathcal{N}_i|$ iterations, and $|\mathcal{N}_i|$ in the worst case is $\mathcal{O}(N-1)$.So the computational
complexity of this algorithm is $\mathcal{O}(CN(N-1))$.

\begin{algorithm}[!t] 
	\caption {Broadcasting Content Selection Algorithm} \label{alg:The selection algorithm}
	\noindent\textbf{Initialization:} Obtain the location of each OBU;
	obtain the neighbor set and possessed content set of each OBU;
	set $\text{Num}=0$; set $T=\infty$; set $\gamma_\text{fin}^i=0$ and $\mathcal{N}_i^*=\emptyset$ of each OBU;
	\begin{algorithmic}[1]
		\FOR{OBU $i$ ($1\leq i\leq N$)}
		 \FOR{possessed content $\gamma_i^p$ ($\gamma_i^p\in\Gamma_i$)}
         \STATE $\text{num}=|\mathcal{N}_{i^*}^{\gamma_i^p}|$;
         \IF{$\text{num}>\text{Num}$}
         \IF{$\text{num}>B_t$}
         \FOR{OBU $j$ ($j\in\mathcal{N}_i$)}
         \STATE calculate $T_{i,j}^{\gamma_i^p}$;
         \ENDFOR
         \STATE select $B_t$ neighbor OBUs $j_1, j_2, ..., j_{|B_t|}\in\mathcal{N}_{i^*}^{\gamma_i^p}$ with relatively small $T_{i,j_1}^{\gamma_i^p}$, $T_{i,j_2}^{\gamma_i^p}$, ..., $T_{i,j_{|B_t|}}^{\gamma_i^p}$;
         \IF{$\text{max}\{T_{i,j_1}^{\gamma_i^p}, T_{i,j_2}^{\gamma_i^p}, ..., T_{i,j_{|B_t|}}^{\gamma_i^p} \}<T$}
         \STATE $\gamma_\text{fin}^i=\gamma_i^p$;
         \ENDIF
         \ELSE
         \STATE $\gamma_\text{fin}^i=\gamma_i^p$;
         \ENDIF
         \ELSE
         \IF{$\text{num}=\text{Num}$}
         \FOR{OBU $j$ ($j\in\mathcal{N}_i$)}
         \STATE calculate $T_{i,j}^{\gamma_i^p}$;
         \ENDFOR
         \IF{$\text{max}\{T_{i,j_1}^{\gamma_i^p}, T_{i,j_2}^{\gamma_i^p}, ..., T_{i,j_{|\mathcal{N}_{i^*}^{\gamma_i^p}|}}^{\gamma_i^p} \}<T$}
         \STATE $\gamma_\text{fin}^i=\gamma_i^p$;
         \ENDIF
         \ENDIF
         \ENDIF
         \ENDFOR
         \STATE Output $\gamma_\text{fin}^i$ and $\mathcal{N}_{i}^*$;
		\ENDFOR
	\end{algorithmic}
\end{algorithm}

Broadcasting contents of different OBUs are decided by this broadcasting content selection algorithm, and then we will construct the coalition formation game.

\section{Full-duplex Coalition Formation Algorithm }\label{S6}
In this section, we propose a FD coalition formation algorithm, and provide some related definitions and some necessary proofs.

\subsection{Coalition Formation Concepts}\label{S6-1}
Since the proposed solution is based on the cooperation of players, we model the PCD problem as a coalition formation game \cite{zhu, The1, The2}. After determining the broadcasting content of each OBU by Algorithm \ref{alg:The selection algorithm}, we need to select OBUs that can broadcast their selected broadcasting contents at the same time. The set of simultaneous broadcasting OBUs is defined by $S$. The limitations of simultaneous transmissions of OBUs have been detailed in Section \ref{S4}. In addition to constrains of simultaneous broadcasting, selecting OBUs into the simultaneous broadcasting group to acquire more contents is the key. This problem is modeled as a coalitional game with a transferable utility, and the definition of the coalitional game is as follows.
\begin{definition}\label{D1}
A coalitional game is defined by a pair $(\mathcal{N},V)$, where $\mathcal{N}$ is the set of game players and $V$ is a function over the real line, $V(S)$ is a real number describing the value that coalition $S\subseteq\mathcal{N}$ can receive, the members of $S$ can be distributed in arbitrary manner, and the coalitional game is with transferable utility.
\end{definition}

For any coalition $S$, the value $V(S)$ comes from cooperations among OBUs in coalition $S$. In order to get the accurate value of $V(S)$, we need to calculate the total revenue generated by coalition $S$ first. The total revenue of $S$ is calculated by utility function $U(S)$.

According to the contribution of content transmissions between OBUs, we define the utility function $U(S)$ is proportional to the number of network's received contents which are from broadcasting OBUs in $S$. A certain number of contents can be received from broadcasting OBUs in $S$, and the set of these contents is denoted by $C_t$. Thus, the utility function $U(S)$ is given by
\begin{equation}\label{eqU}
U(S)=\alpha(\sum_{i\in \mathcal{N}}\sum_{c\in C_t}\delta_i^c).
\end{equation}
where $\alpha>0$ is a utility calculation factor.

The player cooperations can increase the number of received contents of OBUs, but these gains are limited by inherent costs that need to be paid by OBUs for broadcasting and receiving. These costs can be captured by a cost function $C(S)$ which limits the total revenue. Since broadcast durations of OBUs in each coalition are short, we hold that the amount of neighbors of each OBU is approximately stable. Thus, the cost function $C(S)$ is considered as follows
\begin{eqnarray}\label{eq13}
C(S)=
\begin{cases}
\mu|S|,  &\text{if}|S|>1,\\
0, &\text{otherwise},\\
\end{cases}
\end{eqnarray}
where $\mu>0$ is the pricing factor. OBUs in a coalition need to determine broadcast contents, synchronize and transmit contents to other OBUs. Thus, for each coalition $S\in \mathcal{N}$, OBUs need to pay a cost for coordination, which is an increasing function of the coalition size such as in (\ref{eq13}).

Consequently, the value $V(S)$ of any coalition $S$ can be expressed by the utility function in (\ref{eqU}) and the cost function in (\ref{eq13}), which is given by
\begin{equation}\label{eq14}
V(S)=U(S)-C(S).
\end{equation}

This value function quantifies the effective revenue of OBU cooperations in a coalition $S$. In the next section, a coalition formation algorithm will be devised to obtain the coalition with the utility function.

In fact, coalition formation is a very noticeable topic in game theory \cite{22, 23, 24}. Choosing the players to join or leave a coalition based on well-defined preferences is one of the major approaches for forming coalitions. This approach for coalition formation is based on many existing coalition formation concepts, such as the hedonic games or merge-and-split algorithm \cite{23,24,25}. For the proposed coalition formation of the PCD, we introduce some related definitions.

First, the proposed cooperation model entails the formation of disjoint coalitions, which is given as follows
\begin{definition}\label{D2}
A coalitional structure or partition is defined as the coalition set $\Pi=\{S_1,..., S_l\}$, $\Pi$ partitions the OBU set $\mathcal{N}$, coalitions in $\Pi$ are disjoint, i.e., $\forall k$, $S_k\subseteq\mathcal{N}$, $\bigcup_{k=1}^lS_k=\mathcal{N}$ (the OBUs in the same coalition can broadcast simultaneously).
\end{definition}

In order to make explanations more convenient, a related definition is given by
\begin{definition}\label{D3}
For any OBU $i\in\mathcal{N}$, given a partition $\Pi$, the coalition $S_k\in\Pi$, $i\in S_k$, the coalition of $i$ in $\Pi$ is denoted as $S_\Pi(i)$.
\end{definition}

Based on the relation of preferences in the coalition formation game, each OBU must compare and order its potential coalitions. To be specific, each player must select a coalition which it prefers to be a member of. For evaluating these preferences over the coalitions, the definition of preference relation is introduced.
\begin{definition}\label{D4}
For any OBU $i\in\mathcal{N}$, a preference relation $\succeq_i$ is defined as a binary relation over the partition of all coalitions that OBU $i$ can possibly form, it is complete, reflexive, and transitive, i.e., the set \{$S_k\subseteq\mathcal{N}:i\in S_k$\}.
\end{definition}

Hence, given any two coalitions $S_1, S_2\subseteq\mathcal{N}$, $S_1\succeq_iS_2$ implies that OBU $i$ prefers being a member of coalition $S_1$ with $i\in S_1$ over coalition $S_2$ with $i\in S_2$, or at least, OBU $i$ prefers both coalitions equally. On the basis of this preference relation $\succeq_i$, a asymmetric counterpart is denoted by $\succ_i$, $S_1\succ_iS_2$ implies that OBU $i$ strictly prefers being a member of coalition $S_1$ over coalition $S_2$. In order to specifically compare or sort the coalitions, we concrete the preference relation into a function which is used to compare the profit of received contents of different coalitions. To be specific, the profit of the coalition can be calculated with the amount of contents which are successfully received. Therefore, we utilize the value $V(S)$ in \ref{eq14} to compare the profit brought by OBU $i$ joining any coalition. In addition, the individual profits of other members in any coalition cannot be damaged by the joining of OBU $i$. The individual profit of OBU $j$ in coalition $S$ can be calculated with the amount of neighbors that can receive broadcasting content $\gamma_{\text{fin}}^j$ from OBU $j$. Due to collisions of simultaneous broadcasting OBUs, the neighbors that can receive broadcasting content from any OBU in the coalition may change when members of this coalition change. The individual profit of OBU $j$ in coalition $S$ can be expressed as $\psi_S^j$.

Based on the value $V(S)$, we propose the following preference relation ($S_1, S_2\subseteq\mathcal{N}$)
\begin{equation}\label{eq16}
\begin{split}
S_1\succ_iS_2\Leftrightarrow&V(S_1\cup\{i\})>V(S_2\cup\{i\})\\
&\&\psi_{S_1\cup\{i\}}^j\geq \psi_{S_1}^j, \forall j\in S_1.
\end{split}
\end{equation}

These coalition formation concepts have been defined in this section, and then we will propose the coalition formation algorithm over these definitions.

\subsection{Full-duplex Coalition Formation Algorithm}\label{S5-2}
\begin{algorithm}[t!]
\caption{The Full-duplex Coalition Formation Algorithm} \label{alg:The algorithm}
\noindent\textbf{Initialization:} The OBUs in the network are randomly divided into an initial partition $\Pi_{\textit{ini}}$; Set the history collections $h(i)=\emptyset$, $\forall i\in\mathcal{N}$; Set the current partition $\Pi_{\textit{cur}}=\Pi_{\textit{ini}}$; Set the final Nash-stable partition $\Pi_{\textit{fin}}=\Pi_{\textit{ini}}$;
\begin{algorithmic}[1]
\REPEAT
\STATE Randomly choose an OBU $i\in\mathcal{N}$ with current partition $\Pi_{\textit{cur}}$, and denote its current coalition by $S_k=S_\Pi(i)$;
\STATE Search for a coalition $S_m\in\Pi_{\textit{cur}}\cup\emptyset$, where $S_m\cup\{i\}\succ_iS_k$, $S_m\neq S_\Pi(i)$, $S_m\cup\{i\}\not\in h(i)$;
\STATE Perform the broadcasting packet selection algorithm as per \ref{S5}, and re-calculate the individual profit $\psi_S^j$ of any OBU $j$.
\IF {the switch operation from $S_k$ to coalition $S_m\in\Pi_{\textit{cur}}\cup\emptyset$ exists}
\STATE Add the current coalition $S_{\Pi_{cur}}(i)$ to the history collection $h(i)$;
\STATE OBU $i$ leaves the current coalition $S_{\Pi_{cur}}(i)$ and joins the new coalition $S_{\Pi_{new}}(i)$;
\STATE Update the current partition $\Pi_{\textit{cur}}$ to the new partition $\Pi_{\textit{cur}}^\prime$;
$\Pi_{\textit{cur}}=\Pi_{\textit{cur}}^\prime$;
\STATE Since OBU $i$ joins the new coalition $S_{\Pi_{new}}(i)$ and improves its payoff, update the cost of all the coalitions in new partition;
\ENDIF
\UNTIL{the partition converges to a final Nash-stable partition $\Pi_{\textit{fin}}$}
\end{algorithmic}
\end{algorithm}

In order to choose the most appropriate coalition for each OBU, we propose an algorithm that allows the OBU to take distributed decisions for selecting which coalitions to join. The OBUs form disjoint coalitions by switching operations, which is defined as following.
\begin{definition}\label{D5}
Given a partition $\Pi=\{S_1,..., S_l\}$ of the OBUs set $\mathcal{N}$, if any OBU $i\in\mathcal{N}$ decides to leave its current coalition $S_\Pi(i)=S_k$ and join another coalition $S_m\in\Pi\cup\{\emptyset\}$, perform a switch operation from $S_k$ to $S_m$. The partition $\Pi$ ($\Pi\subseteq\mathcal{N}$) is modified into a new partition $\Pi^\prime$ such that $\Pi^\prime=(\Pi\setminus\{S_k, S_m\})\cup\{S_k\setminus\{i\}, S_m\cup\{i\}\}$.
\end{definition}

The basic rule for performing the switch operations is given as follow

\noindent\emph{\textbf{Switch Rule 1}}: Given a partition $\Pi=\{S_1,..., S_l\}$ of the OBUs set $\mathcal{N}$, for any OBU $\forall i\in\mathcal{N}$, if and only if $S_m\cup \{i\}\succ_i S_k$ and $S_m\cup \{i\}\not\in h(i)$ ($S_k\in\Pi, S_m\in\Pi\cup\{\emptyset\}$), a switch operation from $S_k$ to $S_m$ is allowed.

Here $h(i)$ is a history set of coalitions that OBU $i$ has visited in the past and then left. In this rule, any OBU $i\in\mathcal{N}$ can leave its current coalition $S_\Pi(i)$, and join another coalition. The new coalition is strictly preferred over last coalition through the preference relation defined in (\ref{eq16}). The coalition formation game is specifically summarized in Algorithm \ref{alg:The algorithm}. In this algorithm, every OBU try to select its top preferred coalition by performing switch operations. We assume that the order in which the OBUs make their switch operations is random. As each time a switch operation as (\ref{D5}) is performed, the algorithm needs to re-determine the neighbors of each broadcasting OBU and re-perform the broadcasting content selection algorithm as per \ref{S5}. With the partition changing, the best broadcasting contents selected by the broadcasting OBUs may also change. The proposed algorithm needs to adopt the dynamic broadcasting content selection, thereby a more realistic result of utility function comparison can be obtained. Finally, we will get the optimal OBU coalition. The convergence of the proposed coalition formation algorithm is guaranteed as follows

\emph{\textbf{Theorem 1:}}
Whatever the initial partition $\Pi_{ini}$ is, the proposed coalition formation-based algorithm can map to a series of switch operations, and the algorithm will always converge to the final partition $\Pi_{fin}$ which is composed of many disjoint coalitions.

\begin{proof}
For the proof of this theorem, we denote the partition which is formed during turn $k$ of any OBU $i\in\mathcal{N}$ by $\Pi_{n_k}^k$, where $n_k$ is the number of switch operations of the turn $k$. As the OBU is randomly selected of each turn, the partition may or may not change, details are as follows
\begin{equation}\label{eq17}
\Pi_{n_k}^k=\Pi_{n_{k+1}}^{k+1}, n_k=n_{k+1},
\end{equation}
where the number of switch operations is $n_k$ and is unchanged, the meaning of (\ref{eq17}) is that there is no possible switch operation of the turn $k+1$ and the partition does not change.

\begin{equation}\label{eq18}
\Pi_{n_t}^t\rightarrow\Pi_{n_{t+1}}^{t+1}, n_t\neq n_{t+1},
\end{equation}
(\ref{eq18}) means that a switch operation is performed in the turn $t+1$ and the new partition is yielded.

From the preference relation defined in (\ref{eq16}), it can be found that a single switch operation of any OBU $i\in\mathcal{N}$ may lead to yield an unvisited partition or a previously visited partition with a non-cooperative OBU $i$ (OBU $i$ is a singleton coalition of the new partition). If there is a non-cooperative OBU $i$ in partition, it need decide to join a new coalition or remain noncooperative. If OBU $i$ remains non-cooperative, the current partition cannot be changed to any visited partitions in the next turn, as shown in (\ref{eq17}). If OBU $i$ decides to join a new coalition, the switch operation made by OBU $i$ will form an unvisited partition without non-cooperative OBUs, as shown in (\ref{eq18}). No matter how to do it, an unvisited partition will be formed. Besides that, as the well known fact is the number of partitions of a set is given by the Bell number \cite{22}, there are finite different partitions in total. So the entire converging process is given by
\begin{equation}\label{eq19}
\Pi_0^1\rightarrow\Pi_{1}^{2}=\Pi_{1}^{3}\rightarrow...\rightarrow\Pi_{n_s}^{s}.
\end{equation}

Because the different partitions of the fixed OBUs is finite and each switch operation yields an unvisited partition, the number of transformations in (\ref{eq19}) is finite, and the sequence in (\ref{eq19}) will always terminate and converge to a final partition $\Pi_{\textit{fin}} = \Pi_{n_s}^{s}$ after $s$ turns. Thus, the coalition formation of the proposed algorithm will converge to a final network partition $\Pi_{\textit{fin}}$ composed of a number of disjoint OBU coalitions, which completes the proof.
\end{proof}

The proposed algorithm finally converges to the partition $\Pi_{inf}$ and stays in a stable state. The stability of $\Pi_{inf}$ is defined by the following stability concept \cite{23}
\begin{definition}\label{D6}
For any partition $\Pi=\{S_1,..., S_l\}$, if $\forall i\in\mathcal{N}$, $S_\Pi(i)\succeq_iS_k\cup\{i\}$ for all $S_k\in\Pi\cup\{\emptyset\}$, it is Nash-stable.
\end{definition}

The definition \ref{D6} implies that, if there is no OBU has an incentive to perform a switch operation from its current coalition to another coalition, the current partition must be a Nash-stable partition. The Nash-stable partition implies that any OBU $i$ does not prefer to be the part of any other coalition $S_k$ over being the part of its current coalition $S_\Pi(i)$, as per (\ref{eq17}). Besides, OBU $i$ cannot hurt the profits of other OBUs in Nash-stable partition.

\textbf{\emph{Proposition 1:}}
The final partition $\Pi_{\textit{fin}}$ of our proposed coalition formation algorithm is Nash-stable.

\begin{proof}
We make a hypothesis that the final partition $\Pi_{\textit{fin}}$ resulting from the proposed coalition formation algorithm is not Nash-stable. Consequently, there must be an OBU $i\in\mathcal{N}$ and a coalition $S_k$ in partition $\Pi_{\textit{fin}}$, which can satisfy the conditions $S_k\cup\{i\}\succ_i S_{\Pi_{\textit{fin}}}(i)$, and hence, OBU $i$ can perform a switch operation. So this situation contradicts with the fact that $\Pi_{\textit{fin}}$ is the result of the convergence of the Algorithm \ref{alg:The algorithm}. The hypothesis cannot be established. Thus, any partition $\Pi_{\textit{fin}}$ resulting from the coalition formation algorithm is Nash-stable.
\end{proof}

For V2V communications, vehicles exchange their possessed contents to get different popular contents by the proposed coalition formation algorithm, but the coalition formation algorithm is limited by the network scale. If the network scale is large with too many OBUs in the network, contents cannot be shared in time and the delay for receiving more contents is long. In addition, in the lanes where vehicles are sparsely distributed, there may be some vehicles that cannot communicate with other vehicles due to distance and are separated from other vehicles. Thus, the splitting of OBUs is needed to catch up the changes of environment and adapt to the dynamic number of OBUs. Once the number of OBUs exceeds the threshold or some OBUs split into the disconnected parts, all the OBUs in the network automatically split into multiple subnetworks without any overlap. There have been many algorithms of network splitting in the existing literature \cite{tian}, we will not repeat it in this paper. So we can conclude that the number of OBUs involved in our scheme is not large, or OBUs will split into multiple subnetworks and execute the algorithm in each subnetwork.

\section{Simulation Results }\label{S7}

\subsection{Simulation Setup}\label{S7-1}
In this section, we simulate performances of the proposed scheme and other existing schemes. We consider a straight 2-lane highway which has been introduced in Section \ref{S3}. V2V communications are in the 28GHz mmWave vehicle network. To ensure that contents can be transmitted normally when vehicles are moving fast, we update positions of vehicles and calculate new transmission rates of links every 10 slots, and then transmission scheme continues to be executed. The simulation parameters of the highway scenario are listed in Table \ref{tab:parameters}. And we set parameters of mobility model, antenna model, and channel model with reference to \cite{tian}, \cite{V2I}, and \cite{Na3}.

\begin{table}[t!]
 \caption{\label{tab:parameters} \small Simulation Parameters}
 \centering
 \begin{tabular}{lcl}
  \toprule
  Parameter&Symbol&Value\\
  \midrule
  Length of the simulation area&$L$&$3000m$\\
  Number of time slots&$M$&$2000$\\
  Number of OBUs in the network &$N$&$6\sim14$\\
  Slot duration&$\Delta t$&0.1$ms$\\
  Transmit power&$P_t$&30 dBm\\
  System bandwidth&$W$&800MHz\\
  Background noise&$N_0$&-134dbm/MHz\\
  SI cancelation level&$\beta$&$10^{-8}$\\
  SINR threshold of the V2V link &$th_{min}$&$20 dB$\\
  Number of contents&$C$&$20$\\
  Content size &$D$&$50Mb\sim500Mb$\\
  Acceleration&$a$&$1m/s^2$\\
  Probability of changing speed&$p$&$0.1$\\
  Minimal speed&$v_{min}$&$20m/s$\\
  Maximal speed&$v_{max}$&$40m/s$\\
  Minimal distance&$d_{min}$&$50m$\\
  Maximal distance&$d_{max}$&$300m$\\
  Pricing factors&$\alpha, \mu$&$10,1$\\
  Path loss exponent&$\alpha_l$&$2$\\
  Fading factor of channel model&$m$&$2$\\
  Maximum number of V2V beams&$B_t$&$3$\\
  Maximum antenna gain&$G_0$&$20dBi$\\
  Maximum attenuation of antenna model&$A_m$&$26dB$\\
  3dB beam-width in degrees&$\theta_{3dB}$&$17.5^\circ$\\
  \bottomrule
 \end{tabular}
\end{table}

In the simulations, we compare the following performance metrics of each scheme respectively.

1) \emph{Possessed contents}: The number of contents possessed by all OBUs.

2) \emph{Fairness}: Fairness performance is denoted by the jains fairness measure. The Jains fairness measure in \cite{fair} can be used to determine whether the individual profit of each OBU in coalition are fairly achieved. The number of neighbors that can receive the broadcasting content from OBU $i$ ($i\in S$) is $\psi_S^i$. Thus, the Jains fairness measure is
\begin{equation}\label{eq18}
J(\psi_S^1, \psi_S^2, ..., \psi_S^{|S|})=\frac{(\sum_{i\in S}\psi_S^i)^2}{|S|\cdot\sum_{i\in S}(\psi_S^i)^2 }.
\end{equation}

3) \emph{Number of switch operations}: The number of switch operations that are performed until the partition achieves a Nash-stable partition.

4) \emph{CPU time}: CPU time is the actual duration for the entire scheme executing once.

Existing schemes compared with the proposed scheme (\emph{FD cooperative scheme}) are as follows

1) \emph{coalition game scheme}\cite{TITS}: It uses V2V communications to transmit and relay contents in the vehicular network, and is also based on the coalition formation game. The utility function is given based on the minimization of average network delay.

2) \emph{non-cooperative scheme}: Broadcasting content of each OBU is selected randomly, and there is no coordination between OBUs. Broadcasting OBUs transmit contents to all their neighbors without considering collisions.

In order to obtain more reliable average results, all simulations in this paper perform 200 times.

\subsection{Simulation Results} \label{S6-2}
\begin{figure}[t!]
  \begin{center}
  \includegraphics[width=3.3in]{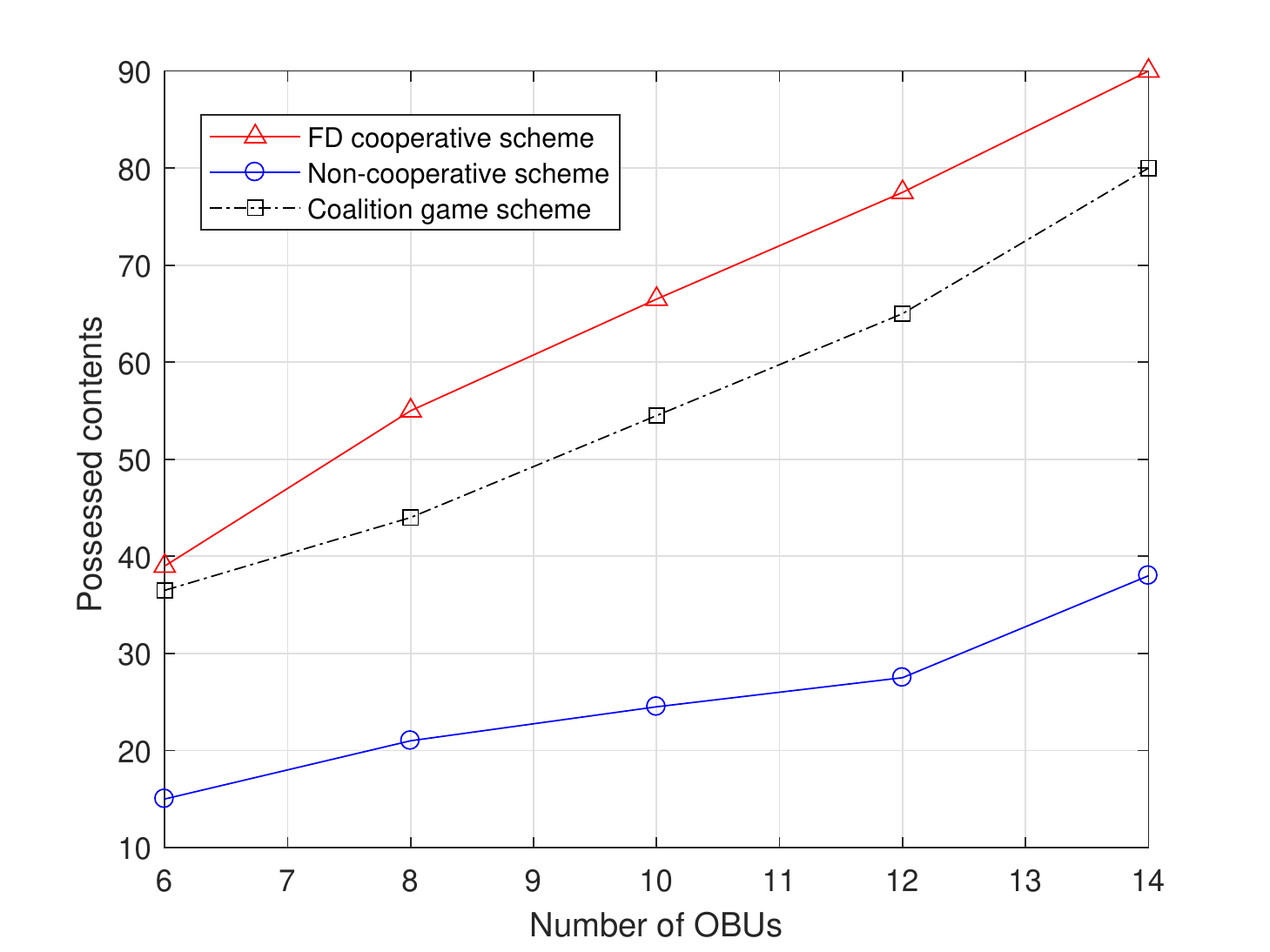}
  \caption{ Number of possessed contents versus the number of OBUs.} \label{fig:con}
  \end{center}
\end{figure}

In Fig. \ref{fig:con}, we plot the number of possessed contents under different OBU amounts. We can see that possessed content amounts of three schemes are all increasing with the increased number of OBUs. The more OBUs in vehicle networks, the more OBUs can simultaneously broadcast and the more neighbors can receive contents from broadcasting OBUs. For the non-cooperative scheme, possessed contents are much less that of the proposed cooperative scheme. Besides, the rising trend of the non-cooperative scheme is also more slowly. The non-cooperative scheme randomly selects broadcasting contents of OBUs, and there are so many collisions among broadcasting OBUs. Therefore, the non-cooperative scheme has the relatively worst performance on possessed contents among three schemes. For the coalition game scheme, the growth trend is similar to that of the proposed scheme. Possessed contents of the coalition game scheme are fewer than that of the proposed scheme, since this scheme is proposed to minimize the network transmission delay. When the number of OBUs is 14, the proposed FD cooperative scheme promotes the number of possessed contents by 12.5\% compared with the coalition game scheme, and by 136.8\% compared with the non-cooperative scheme.

\begin{figure}[t!]
  \begin{center}
  \includegraphics[width=3.3in]{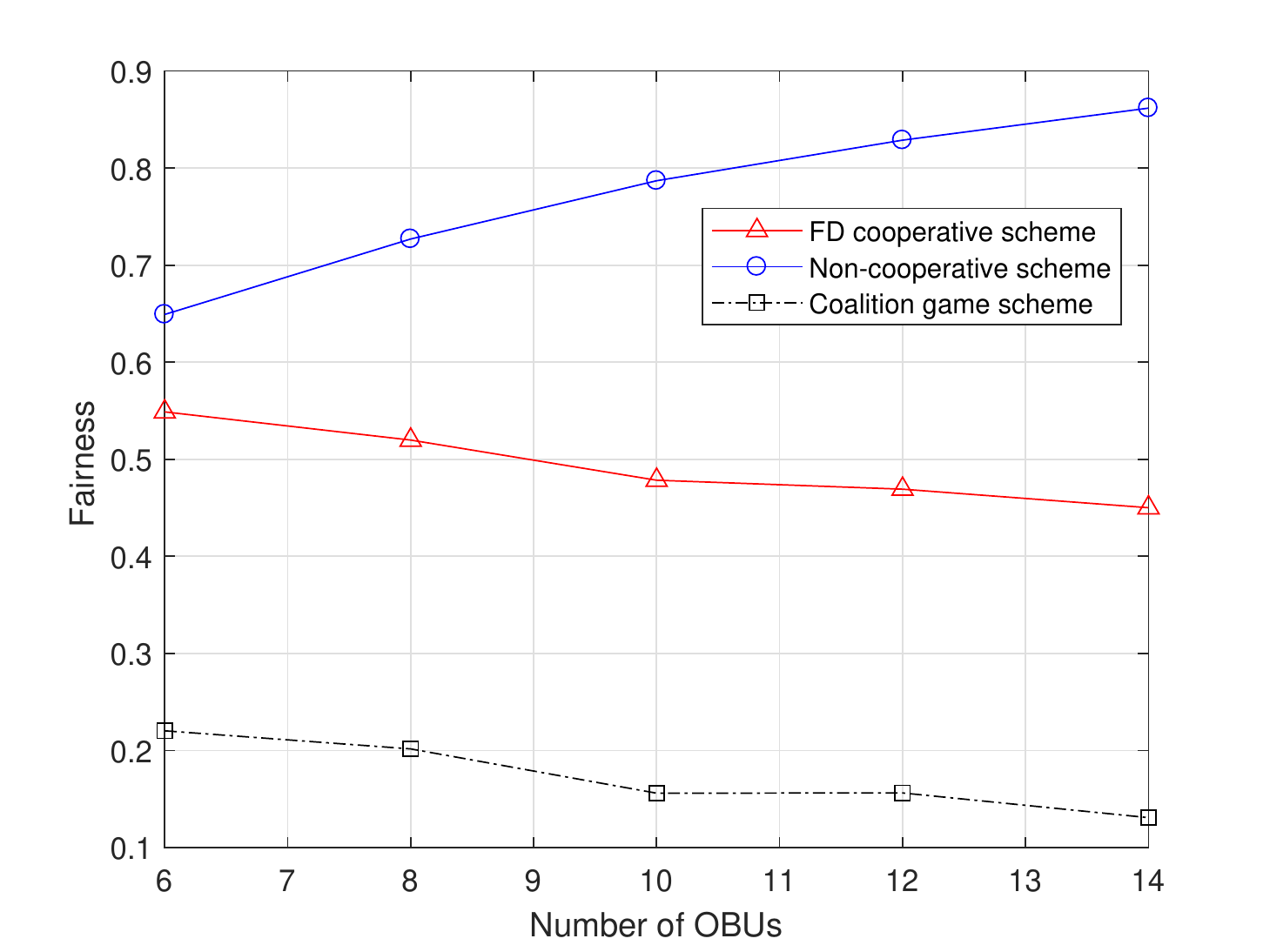}
  \caption{ Fairness under versus the number of OBUs.} \label{fig:fair}
  \end{center}
\end{figure}

In Fig. \ref{fig:fair}, we plot the fairness under different numbers of OBUs. Along with the increase of OBUs, trends of the proposed scheme and the coalition game scheme are both falling, but the trend of the non-cooperative scheme is rising. For the non-cooperative scheme, collisions from no coordination among OBUs leads to few OBUs in coalitions. Few OBUs of each coalition can improve the fairness. With the increased number of OBUs, collisions among OBUs are becoming more, and the number of OBUs that can simultaneously broadcast are decreasing. Thus, the non-cooperative scheme has a rising trend of the fairness. For other two cooperative scheme, coordinations make more OBUs simultaneously broadcast with the increase of OBUs. And then their trends of the fairness both fall due to more broadcasting OBUs. Compare with the coalition game scheme, our proposed scheme protects the individual profit of each OBU in coalition. Therefore, the performance of the proposed scheme on fairness is better than that of the coalition game scheme.

\begin{figure}[t!]
  \begin{center}
  \includegraphics[width=3.3in]{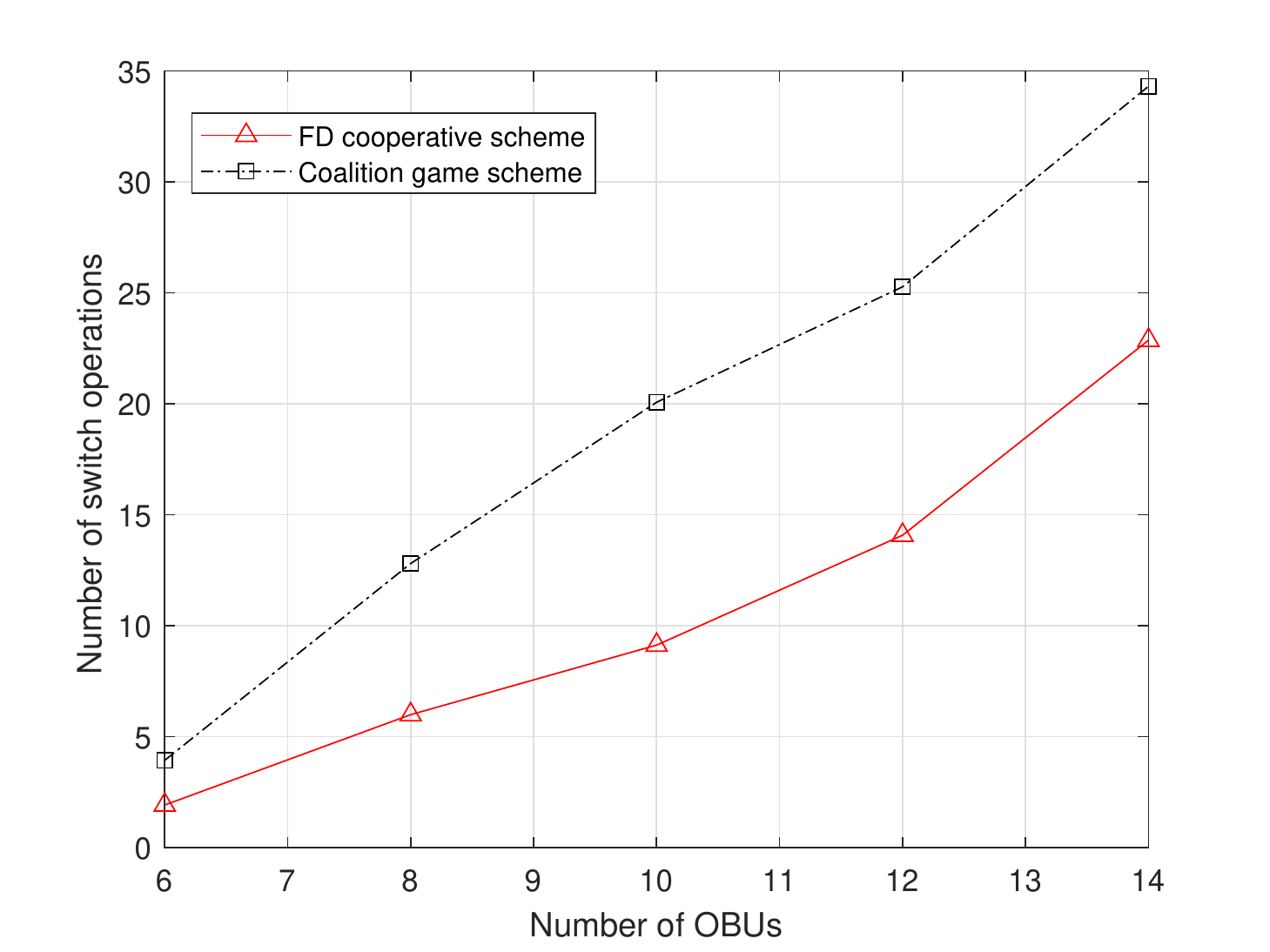}
  \caption{ Number of switch versus the number of OBUs.} \label{fig:switch}
  \end{center}
\end{figure}

In Fig. \ref{fig:switch}, we plot the number of switch operations under different numbers of OBUs. The non-cooperative scheme has no switch operations, so there are only two result curves. The number of switch operations can reflect the complexity of each algorithm. We can see that switch operations of two schemes are both becoming more with the increased number of OBUs. Besides, switch operations of the proposed FD cooperative scheme are less than that of the coalition game scheme. The coalition game scheme seeks to the minimization of average network delay, so it needs to execute coalition game algorithm more times in the fixed time compared with the proposed algorithm. However, Fig. \ref{fig:switch} can illustrate that the complexity of the proposed algorithm is lower.

\begin{figure}[t!]
  \begin{center}
  \includegraphics[width=3.3in]{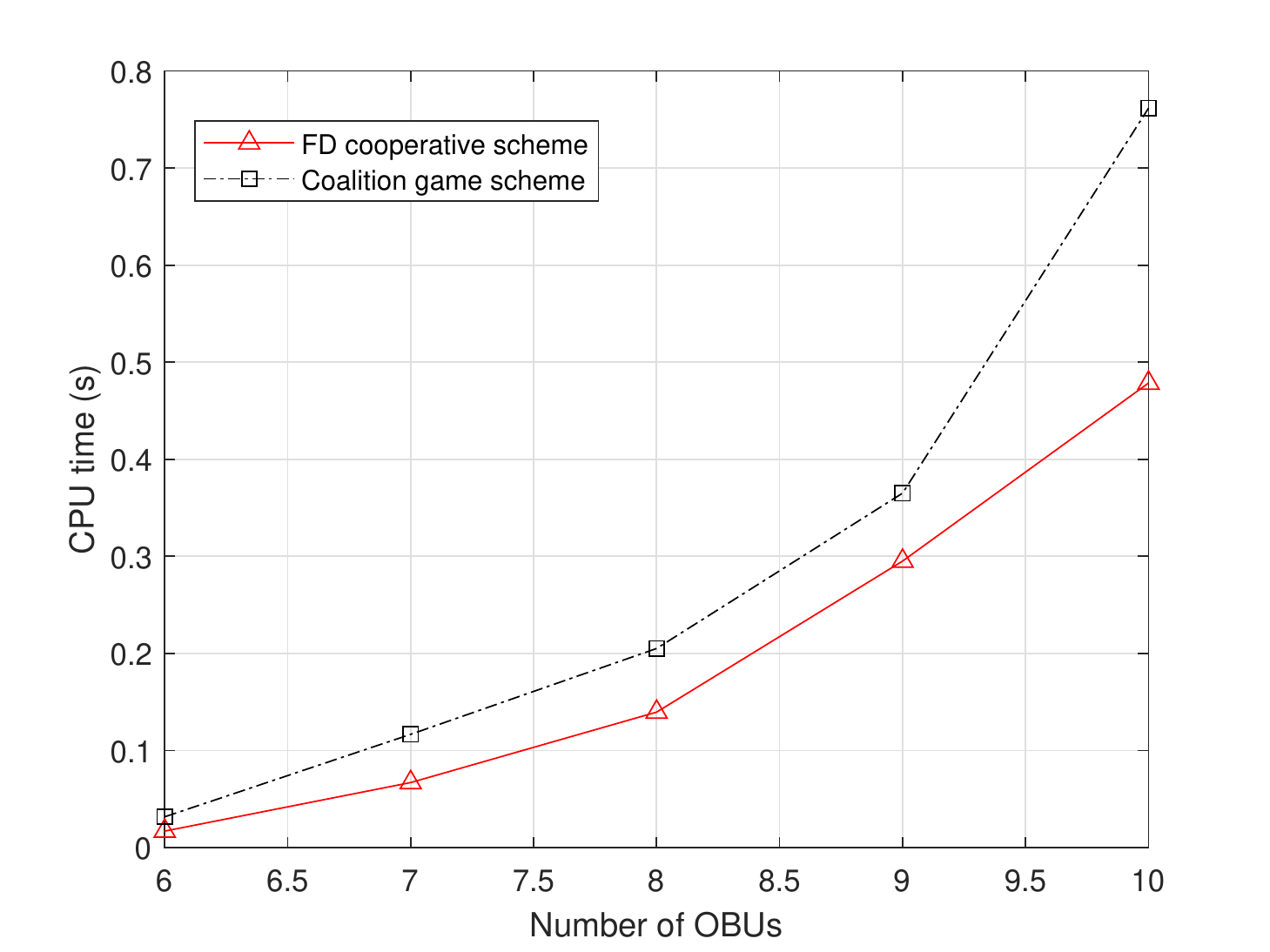}
  \caption{ CPU time versus the number of OBUs.} \label{fig:cpu}
  \end{center}
\end{figure}

In Fig. \ref{fig:cpu}, we plot the CPU time with the different numbers of OBUs. In order to verify the complexity of the entire scheme (including the broadcasting content selection algorithm in Section \ref{S5}), the CPU times of different schemes are calculated. The result is similar to the result in Fig. \ref{fig:switch}, our proposed algorithm has shorter CPU time and lower complexity. And we can also see that the CPU time for executing the entire proposed scheme once is less than a second when the number of OBUs is less than 10.

\begin{figure}[t!]
  \begin{center}
  \includegraphics[width=3.3in]{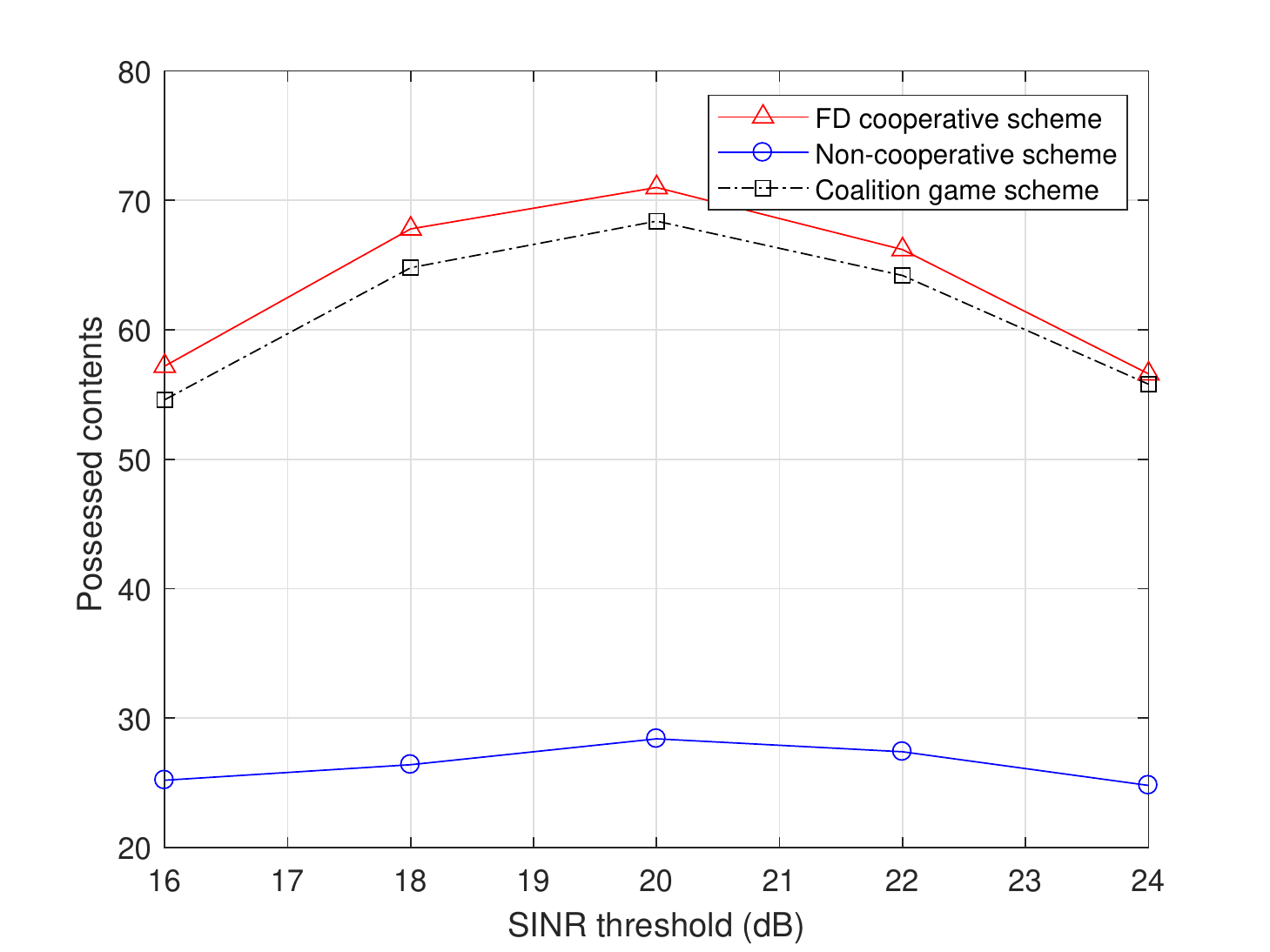}
  \caption{ Number of possessed contents versus the SINR threshold.} \label{fig:thr}
  \end{center}
\end{figure}

In Fig. \ref{fig:thr}, we plot the number of possessed contents under different SINR thresholds. The number of OBUs is set to 10. For these three schemes, overall trends of their result curves are the same. With the increase of SINR threshold, the possessed content numbers of schemes all increase first and then decrease. Properly increasing of SINR threshold can avoid some transmissions with very small rates. And then rates of other transmissions can be increased, which will lead to shorter transmit time and more possessed contents. When the SINR threshold is increased to a certain value (i.e., $th_{min}=20$dB in Fig. \ref{fig:thr}), the numbers of possessed contents of all schemes start to decrease. This is because that too small SINR thresholds weaken the advantage of spatial reuse and reduce possessed contents of the network. So an appropriate actual SINR threshold is important for transmission performances. In other simulations of this paper, we set the value of the SINR threshold to be 20dB.

\begin{figure}[t!]
  \begin{center}
  \includegraphics[width=3.3in]{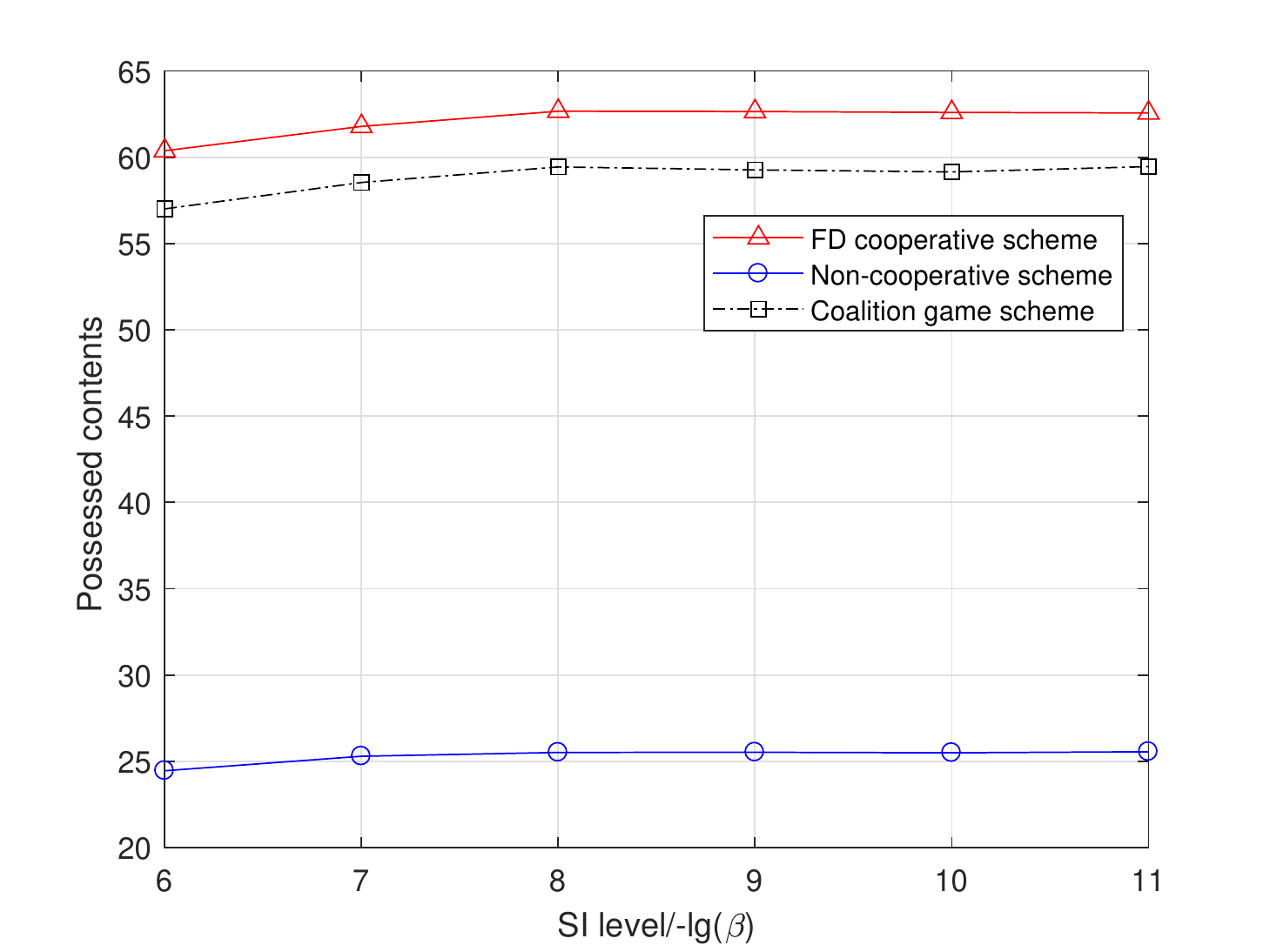}
  \caption{ Number of possessed contents versus the SI level.} \label{fig:beta}
  \end{center}
\end{figure}

In Fig. \ref{fig:beta}, we plot the number of possessed contents under different SI cancelation levels. The number of OBUs is set to 10. The abscissa $x$ is $-lg(\beta)$, i.e., the SI cancelation level $\beta=10^{-x}$. The small value of $\beta$ represents the high SI cancelation level. With the improvement of SI cancelation level, the numbers of possessed contents of three schemes all increase first and then keep stable. Properly increasing of the SI cancelation level can improve transmission rates of broadcasting contents. To some extent, high rates lead to more possessed contents. When the SI cancelation level achieves a certain value (i.e., $\beta=10^{-8}$ in Fig. \ref{fig:beta}), the numbers of possessed contents of three schemes start to be unchanged. In this case, the performance of these schemes mainly depends on different algorithms, and FD communications can no longer bring more promotion.

\section{Conclusion}\label{S8} 
In this paper, we consider the PCD problem in mmWave vehicular networks. We investigate the content transmission by V2V communications and propose the FD cooperative scheme based on coalition formation game. The utility function of the proposed scheme is provided based on the maximization of the received content number. In addition to maximization of the utility function, the proposed scheme ensures the individual profit and related fairness of each broadcasting OBU. Simulation results show that our proposed scheme has much better performances on the number of possessed contents and fairness. Moreover, the proposed scheme has the lower complexity compared with other schemes. In the future work, we will consider a more realistic scene and traffic model. Besides, we will do the performance verification of our algorithm on the actual system platform to further demonstrate the practicality of our scheme.

\end{document}